# Conditional loss probabilities for systems of economic agents sharing light-tailed claims with analysis of portfolio diversification benefits

Claudia Klüppelberg[1] and Miriam Isabel Seifert[2]

December 21, 2016

**Abstract**

We analyze systems of agents sharing light-tailed risky claims issued by different financial objects. Assuming exponentially distributed claims, we obtain that both agents' and system's losses follow generalized exponential mixture distributions. We show that this leads to qualitatively different results on individual and system risks compared to heavy-tailed claims previously studied in the literature. By deducing conditional loss distributions we investigate the impact of stress situations on agents' and system's losses. Moreover, we present a criterion for agents to decide whether holding few objects or portfolio diversification minimizes their risks in system crisis situations.

[1] Technische Universität München, Boltzmannstraße 3, 85748 Garching, Germany. Email: cklu@tum.de

[2] Ruhr-Universität Bochum, Universitätsstraße 150, 44801 Bochum, Germany. Email: miriam.seifert@rub.de

# 1 Introduction

By monitoring a financial system a CEO (regulator) should assess risk exposures of business lines or single companies in order to determine capital reserves required in case of unexpected large losses. A regulator's assessment requires the following information: What are the individual agent's risks for large exposures? Which are the dominant objects able to cause serious losses to agents or to the entire system? How does the system structure affect the relationship between individual risks and the system risk? We contribute to the current literature by exploring all these issues for light-tailed exponentially distributed losses, which are of particular relevance for standard insurance claims and further related applications (cf. Embrechts et al. [5]).

We consider a system of risky objects which are hold by economic agents forming a financial network. This setting can be illustrated by a bipartite graph of agent-object relationships as shown in Figure 1.1. In this paper, we focus primarily on agents holding various claims in their portfolios and the corresponding risks. As many insurance companies and investment funds hold portfolios with risky claims which may cause critical losses, these issues should be studied in a network context. In particular, it is of importance to investigate consequences of losses for both individual agents and the system as well as to compute the corresponding conditional and unconditional loss probabilities. Such results are of special interest for various types of regulating authorities, which should facilitate stability monitoring for both the system and agents. In this context the analysis of portfolio diversification benefits is of relevance for quantifying individual and system risks; cf. Geluk et al. [8], Ibragimov [10], Mainik and Rüschendorf [20], Pérignon and Smith [25].

The effects of risk aggregation and risk sharing have been studied in the current literature mainly for heavy-tailed claims with a power decay in their tails; see Embrechts et al. [6], Kley et al. [15], Ly Vath et al. [19], Lin et al. [18], Xia [26] among others. Light-tailed distributions, however, provide a suitable description of many risks faced by financial institutions or insurance companies, as it is shown e.g. by Asmussen and Albrecher [2], Hernández and Junca [9], Kaas et al. [14], Kyprianou [17] for the insurance context and by Andersen et al. [1] with examples in the financial context. In particular, the exponential distribution is rather convenient as it allows for the derivation of important results in an explicit form. Moreover, it can often be extended to more general distribution families, such as referring to the Gumbel max-domain of attraction (cf. Mitra and Resnick [24]) or to the Cramér-Lundberg class for ruin estimates (cf. Asmussen and Albrecher [2]).

Currently, there are only a few stochastic results on aggregating and sharing losses with light-tailed distributions. Jiang and Tang [13] study the asymptotic behavior of reinsured losses in a setting of independently and identically distributed exponential claims. Since in their paper all claims have the same parameter $\lambda$, the aggregated claim (system risk) follows an Erlang distribution. Mitra and Resnick [24] analyze the aggregation $X + Y$ for claims $X, Y$ with tail-equivalent distributions in the Gumbel max-domain of attraction, which contains the exponential, Gaussian, and log-normal distributions as specific examples. They assume a certain dependence structure between $X$ and $Y$, which results in their asymptotic independence. Farkas and Hashorva [7] consider two portfolios of Gaussian-like risks and derive limit results for the distribution of portfolio losses. The asymptotic results for optimal allocation problems for exponential claims are obtained in Maume-Deschamps et al. [22] as the capital size goes to infinity. However, none of these papers consider consequences of risk sharing in the context of network or system risks.

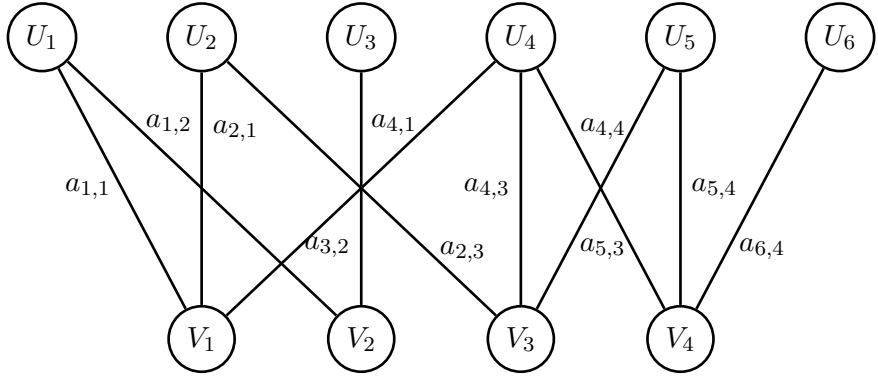

**Figure 1.1:** The risk sharing structure as a bipartite graph for a system of 4 objects with claims $V_j$ hold by 6 agents with exposures $U_i$. The portfolio weights are $a_{i,j}$ for agents $i \in \{1, \ldots, 6\}$ and objects $j \in \{1, \ldots, 4\}$.

In this paper we contribute to the current research by studying independent exponential claims on risky objects with distinct tail decays in the context of risk sharing with a bipartite graph representation. Our framework is rather general and flexible. We allow for different risk classes modelled by distinct exponential parameters which is a generalization compared to Jiang and Tang [13], Kley et al. [15, 16] and Mitra and Resnick [24], who assume tail-equivalent risks for either light- or heavy-tailed claims. In our setting, the agents may form their portfolios by holding selected claims with different proportions. This implies that the agents' risks as well as the risk of the whole system follow "generalized exponential mixture" distributions; cf. Jasiulewicz and Kordecki [11] and Jewell [12].

Our analysis provides novel results concerning conditional survival functions of the type $P(L_1 > l_1 \mid L_2 > l_2)$ where $L_i$, $i = 1, 2$ denote losses either of specific objects, or of individual agents, or of the entire system. Thereby, we obtain statements both for the exact conditional distributions with finite threshold values $l_i$ and for the asymptotic tail behavior with $l_i \to \infty$. That is, we provide expressions for unconditional and conditional tail probabilities as they contain the information about extreme situations which are of most importance in practice. As the class of generalized exponential mixture distributions has not been investigated with respect to their conditional distributions up to now, our findings contribute to their statistical theory. Our results also could be used for computing standard risk measures like Value-at-Risk and Expected Shortfall.

We start our discussion by deriving the survival functions for both single agent's and system's losses. The important result is given in Theorem 3.5 where we show that the *dominant* impact on both individual and system risks is determined by single distinct objects and that generically the risk-dominant objects for the system do not coincide with those for individual agents. We provide an economic explanation for this interesting phenomenon in Remark 3.6.

In Theorems 4.1 and 4.6 we deduce both exact and asymptotic results for the conditional distributions of agents' and system's losses given that an object claim exceeds a certain threshold value. We show that these conditional distributions do not depend on which object causes this large claim, which is of importance for applications. Both mathematical and economic interpretations of our results are provided in Remarks 4.2 and 4.7, respectively. There we point out that the impact of a large claim on individual or system's losses can be completely quantified solely by their marginal distributions, which implies a substantial policy simplification from the regulator's point of view.

Furthermore, we analyze interdependencies between individual and system risks, in particular for agents with different degrees of portfolio diversification; our results of Theorem 5.1, Proposition 5.2, and Corollaries 5.3 – 5.6 are interpreted in Remark 5.7. Moreover, we introduce in Theorem 5.8 a criterion to decide in view of systemic crisis situations, when it is beneficial for agents to concentrate on a few objects or to diversify.

Finally, we provide a detailed comparison of our results for systems with exponential claims to the corresponding results in heavy-tailed settings. We point out the substantial differences between these two settings; e.g., we show that diversification works rather different for exponential claims compared to heavy-tailed (Pareto) claims.

Our paper is organized as follows. After describing our framework with relevant assumptions in Section 2, we derive in Section 3 the marginal distributions of individual risks as well as of the system risk. In Section 4 we study the impact of large claims on the system's and the agents' losses. In Section 5.1 we analyze the interrelation between individual and system risks, which is of a special importance for system regulation. In Section 5.2 we discuss diversification effects for agents' portfolios in the context of systemic crises. In Section 6 we summarize our findings and compare our results for light tails with those established for heavy-tailed models. The proofs of Theorem 4.1 and 5.1 are postponed to Section 7, whereas the other proofs are placed immediately after the statements.

## 2 Model framework: notions and notations

In this section we formalize the framework for our investigation. We analyze systems which consist of $d$ objects and $n$ agents for some positive integers $d$ and $n$; the objects $j \in \underline{d} := \{1, \ldots, d\}$ cause *claims of size* $V_j > 0$ which are shared among the agents such that the portfolio loss

$$U_i = \sum_{j=1}^{d} a_{i,j} V_j, \quad i \in \underline{n} := \{1, \ldots, n\}, \tag{2.1}$$

expresses the *risk exposure of agent i*. Moreover, the system's loss is denoted as

$$S = \sum_{j=1}^{d} V_j. \tag{2.2}$$

The *portfolio weights* are collected into $(n \times d)$-dimensional matrix $A = (a_{i,j})_{i \in \underline{n}, j \in \underline{d}}$, which is the weighted adjacency matrix to the bipartite graph. Each component of $A$ as well as the column-sums of $A$ have to be less or equal to 1; i.e.,

$$0 \leq a_{i,j} \leq 1 \text{ for all } i \in \underline{n},\, j \in \underline{d}; \quad \sum_{i \in \underline{n}} a_{i,j} \leq 1 \text{ for all } j \in \underline{d}. \tag{2.3}$$

The boundary value 1 corresponds to the case that the risk for the object $j$ is covered in total. We assume the matrix $A$ and, consequently, the bipartite graph to be deterministic.

Throughout this paper, we meet the following two assumptions:

**Assumption 1** The claims $V_j$ are stochastically independent and exponentially distributed with parameters $\tilde{\lambda}_j > 0$ for $j \in \underline{d}$. Hence, the claim $V_j$ has the density $f_{V_j}(x) = \tilde{\lambda}_j \exp(-\tilde{\lambda}_j x)$ for $x > 0$.

Further, let

$$M_{(i)} := \{j \in \underline{d} \mid a_{i,j} > 0\} \tag{2.4}$$

be the set of indices from all objects selected by agent $i \in \underline{n}$, then $U_i$ can be represented as the sum $\sum_{j \in M_{(i)}} X_{i,j}$ of independent exponentially distributed $X_{i,j}$ with parameters

$$\lambda_{i,j} := \tilde{\lambda}_j / a_{i,j}. \tag{2.5}$$

**Assumption 2** For all $i \in \underline{n}$, $k, j \in \underline{d}$ with $k \neq j$ we require that $\tilde{\lambda}_k \neq \tilde{\lambda}_j$ and $\lambda_{i,k} \neq \lambda_{i,j}$.

This assumption ensures that the parameters of the exponential distributions are pairwise distinct, both of the claims $V_j$ and of the weighted claims $a_{i,j} V_j$. This assumption is commonly met for the analysis of generalized exponential mixtures (see e.g. Bergel and Egídio dos Reis [3], McLachlan [23]). In Remark 3.7(ii) we show how to handle the case where the restriction for pairwise distinct parameters is removed.

**Notations and conventions.** Two functions $f$ and $g$ are said to be *asymptotically equivalent* (we write $f \sim g$) if $f(x)/g(x) \to 1$ for $x \to \infty$ and *proportional* (we write $f \propto g$) if $f(x)/g(x) = c$ for some constant $c > 0$. Further, we denote by $|M|$ the cardinality of a set $M$.

The smallest parameters will play an important role for the risk-dominant terms, so that we define their values and the corresponding indices as

$$\ell(i) := \lambda_{i,m(i)} := \min_{j \in M_{(i)}} \lambda_{i,j}, \tag{2.6}$$

$$\tilde{\ell} := \tilde{\lambda}_{\tilde{m}} := \min_{j \in \underline{d}} \tilde{\lambda}_j. \tag{2.7}$$

For a reader's convenience, we distinguish quantities corresponding to the system from quantities corresponding to agents by a tilde. The following table summarizes our notation.

|  | system | agent | definition |
|---|---|---|---|
| parameters: | $\tilde{\lambda}_j$, $\tilde{\ell}$ | $\lambda_{i,j}$, $\ell(i)$ | Assumpt. 1, Eqs. (2.5) – (2.7) |
| object indices (subset of $\underline{d}$): | $\tilde{m}$ | $m(i)$, $M_{(i)}$ | Eqs. (2.4), (2.6), (2.7) |
| mixing proportions: | $\tilde{\pi}_j$ | $\pi_{i,j}$ | Eqs. (3.2), (3.5) below |

## 3  Individual and system risks

In this section we investigate the unconditional distributions of individual agents' risks as well as that of the system risk. After providing the exact results, we deduce their characteristic tail behavior.

Consider an arbitrary agent $i$ which selects objects $V_j$ for $j \in M_{(i)}$ as defined in (2.1) and (2.4).

**Proposition 3.1.** *Under Assumptions 1 and 2, the exposure $U_i$ of agent $i \in \underline{n}$ has density*

$$f_{U_i}(x) = \sum_{j \in M_{(i)}} \pi_{i,j} \lambda_{i,j} \exp(-\lambda_{i,j} x), \quad x > 0, \qquad (3.1)$$

*with mixing proportions*

$$\pi_{i,j} := \begin{cases} \prod_{k \in M_{(i)} \setminus \{j\}} \dfrac{\lambda_{i,k}}{\lambda_{i,k} - \lambda_{i,j}} & \text{for } |M_{(i)}| > 1, \\ 1 & \text{otherwise}. \end{cases} \qquad (3.2)$$

The distributional form (3.1) can be deduced analogously to Jasiulewicz and Kordecki [11, Th. 1] by applying Laplace transformations. From integration of (3.1) it follows that the mixing proportions sum up to one:

$$\sum_{j \in M_{(i)}} \pi_{i,j} = 1, \qquad (3.3)$$

where $\lfloor |M_{(i)}|/2 \rfloor$ of the $|M_{(i)}|$ mixing proportions are negative. Then the exposure $U_i$ with density (3.1) is said to follow a *generalized exponential mixture* (GEM) distribution; see Mathai [21]. This distribution class is also known in the literature as "generalized Erlang", see e.g. Bergel and Egídio dos Reis [3].

**Remark 3.2.** The mixing proportions alternate in sign: If the weighted claims $X_{i,j} = a_{i,j} V_j$ are put (without loss of generality) in ascended order, i.e. $\lambda_{i,1} < \lambda_{i,2} < \cdots < \lambda_{i,|M_{(i)}|}$, then $\pi_{i,j}$ is positive for uneven indices $j$ and negative for even indices, as $\pi_{i,j}$ has exactly $(j-1)$ negative factors in the denominator of (3.2). $\diamondsuit$

An illustration is given by the following simple example.

**Example 3.3.** Let some agent $i$ have exactly two objects 1 and 2 in the portfolio. With $X_{i,j} := a_{i,j} V_j$ we obtain:

$$
\begin{aligned}
f_{U_i}(x) &= \int_0^x f_{X_{i,1}}(x-u) f_{X_{i,2}}(u) \mathrm{d}u \\
&= \lambda_{i,1} \lambda_{i,2} \exp(-\lambda_{i,1} x) \int_0^x \exp((\lambda_{i,1} - \lambda_{i,2})u) \mathrm{d}u \\
&= \frac{\lambda_{i,2}}{\lambda_{i,2} - \lambda_{i,1}} f_{X_{i,1}}(x) - \frac{\lambda_{i,1}}{\lambda_{i,2} - \lambda_{i,1}} f_{X_{i,2}}(x) = \sum_{j=1}^2 \pi_{i,j} f_{X_{i,j}}(x), \quad x > 0,
\end{aligned}
$$

hence, one of the mixing proportions is positive and the other one is negative. $\diamondsuit$

Next we consider the risk of the entire system, which is defined as the aggregation of all claims $S = \sum_{j=1}^d V_j$. It follows a GEM distribution as well:

**Corollary 3.4.** *Under Assumptions 1 and 2, the system risk $S$ has density:*

$$
f_S(x) = \sum_{j=1}^d \tilde{\pi}_j f_{V_j}(x) = \sum_{j=1}^d \tilde{\pi}_j \tilde{\lambda}_j \exp(-\tilde{\lambda}_j x), \quad x > 0, \tag{3.4}
$$

*with mixing proportions*

$$
\tilde{\pi}_j := \begin{cases} \displaystyle\prod_{k \in \underline{d} \setminus \{j\}} \frac{\tilde{\lambda}_k}{\tilde{\lambda}_k - \tilde{\lambda}_j} & \text{for } d > 1, \\ 1 & \text{otherwise}. \end{cases} \tag{3.5}
$$

These mixing proportions $\tilde{\pi}_j$ satisfy properties analogously to those of $\pi_{i,j}$ for individual agents as described above.

In the following theorem we state an important result characterizing our framework with exponential claims. It gives the asymptotic behavior of individual and system risks for large losses and points out their different tail behavior. More precisely, it shows that the claim with the smallest parameter, see (2.6) and (2.7), determines the asymptotics:

**Theorem 3.5.** *Under Assumptions 1 and 2 the survival functions satisfy:*

(i) *for individual risk of agent $i \in \underline{n}$ :*

$$
\begin{aligned}
P(U_i > x) &= \sum_{j \in M_{(i)}} \pi_{i,j} \exp(-\lambda_{i,j} x), \quad x > 0, \\
&\sim \pi_{i,m(i)} \exp(-\ell(i) x) \propto P(a_{i,m(i)} V_{m(i)} > x) \text{ for } x \to \infty;
\end{aligned}
$$

(ii) *for system risk:*

$$P(S > x) = \sum_{j=1}^{d} \tilde{\pi}_j P(V_j > x) = \sum_{j=1}^{d} \tilde{\pi}_j \exp(-\tilde{\lambda}_j x), \quad x > 0,$$
$$\sim \tilde{\pi}_{\tilde{m}} \exp(-\tilde{\ell} x) \propto P(V_{\tilde{m}} > x) \text{ for } x \to \infty.$$

Theorem 3.5 is interpreted in the following Remark 3.6 and illustrated in Figure 3.1.

**Remark 3.6.**

(i) Both system risk $S$ and individual risk $U_i$ of agent $i$ have asymptotically exponential tails, but with different tail decays. The survival function of the system risk is asymptotically proportional to that of the claim $V_{\tilde{m}}$ with the smallest value of the parameters $\tilde{\lambda}_j$, i.e. $\tilde{m} = k$ for $\tilde{\lambda}_k = \min(\tilde{\lambda}_j)$. In contrast, the survival function of the individual risk $U_i$ is determined asymptotically by the claim which – dependent on parameters $\tilde{\lambda}_j$ and scaled by the portfolio weights $a_{i,j}$ – takes the minimal value among all $\lambda_{i,j} = \tilde{\lambda}_j/a_{i,j}$ for $j \in M_{(i)}$ (all claims in the portfolio of agent $i$).

(ii) Due to the different tail decays, generically the individual risks are asymptotically negligible with respect to the system risk:

$$P(U_i > x) = o(P(S > x)) \quad \text{for } x \to \infty$$

for all $i \in \underline{n}$, except for the special case that agent $i$ selects the most risky object $j$ alone; i.e., when $a_{i,\tilde{m}} = 1$.

$\diamondsuit$

**Remark 3.7.** Here we comment on the case of dropping Assumption 2, which implies a possibility of equal parameters for some (or all) claims. Then the following would hold:

(i) The parameters $\lambda_{i,j}$ may coincide for different $j$-values and, hence, the distribution of individual exposure $U_i$ is not GEM any longer. If – as a special case – all $\lambda_{i,j}$ for $j \in M_{(i)}$ coincide, then $U_i$ is Erlang distributed. In general, if only some $\lambda_{i,j}$ coincide, $U_i$ follows a *generalized Erlang mixture* distribution, cf. Jasiulewicz and Kordecki [11]. This applies for the system's loss $S$ as well.

(ii) The corresponding results of Proposition 3.1 and Corollary 3.4 for $\lambda_{i,k} = \lambda_{i,j}$ (or $\tilde{\lambda}_k = \tilde{\lambda}_j$) can be obtained as limits $\lambda_{i,k} \to \lambda_{i,j}$ (or $\tilde{\lambda}_k \to \tilde{\lambda}_j$). Consider

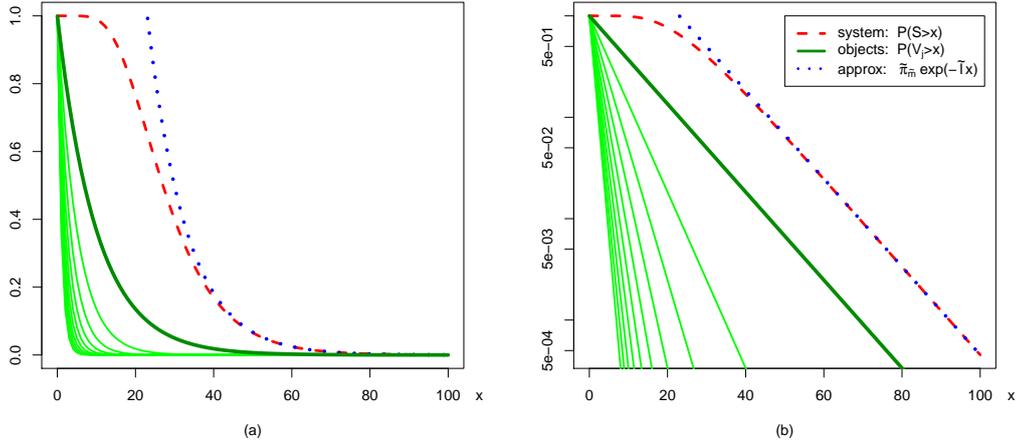

**Figure 3.1:** (a): Survival functions for $d = 10$ objects having exponentially distributed claim sizes with equidistant parameters $\tilde{\lambda}_1 = 0.1, \tilde{\lambda}_2 = 0.2, \ldots, \tilde{\lambda}_{10} = 1.0$ (solid lines), and the function $x \mapsto \tilde{\pi}_{\tilde{m}} \exp(-\tilde{\ell} x)$ (dotted line) as an approximation of system $P(S > x)$ (dashed line); cf. Theorem 3.5.
(b): Same functions as in (a) with a logarithmic $y$-axis to illustrate that parameter $\tilde{l}$ of the approximating function coincides with parameter $\tilde{\lambda}_1$ of the dominant object claim $V_{\tilde{m}} = V_1$ (dotted line and bold solid line).

Example 3.3, where taking such limit leads to an Erlang distribution:

$$\begin{aligned}
\lim_{\lambda_{i,2} \to \lambda_{i,1}} P(U_i > x) &= \lim_{\lambda_{i,2} \to \lambda_{i,1}} \frac{\lambda_{i,2} \exp(-\lambda_{i,1} x) - \lambda_{i,1} \exp(-\lambda_{i,2} x)}{\lambda_{i,2} - \lambda_{i,1}} \\
&= \lim_{\lambda_{i,2} \to \lambda_{i,1}} \left( \exp(-\lambda_{i,1} x) + \lambda_{i,1} x \exp(-\lambda_{i,2} x) \right) \\
&= (1 + \lambda_{i,1} x) \exp(-\lambda_{i,1} x), \quad x > 0.
\end{aligned}$$

The asymptotic results in Theorem 3.5 give Erlang tails of order $q$, where $q$ is the number of *asymptotically dominant claims* with the smallest parameters; i.e., for statements (i) and (ii) we have $q = |\{j \in M_{(i)} \mid \lambda_{i,j} = \min_{k \in M_{(i)}} \lambda_{i,k}\}|$ or $q = |\{j \in \underline{d} \mid \tilde{\lambda}_j = \min_{k \in \underline{d}} \tilde{\lambda}_k\}|$, respectively. $\diamondsuit$

# 4 Conditional results on agents' and system losses given large claims

In this section we investigate the conditional distributions for the system risk $S$ and for an arbitrary agent's risk $U_i$ given that an object causes a large claim. More precisely, we derive the conditional survival functions and densities of $S$ and $U_i$ given that a claim $V_j$ exceeds a certain threshold and, conversely, those of $V_j$ given that $S$ or $U_i$ exceeds some threshold.

In the following theorem we deduce a characteristic property which illustrates the influence of a large object's loss on the system risk. Hereafter we exclude trivial cases where the corresponding conditional probability is equal to 1, e.g. in the following theorem we only consider $s > v$ and $\alpha > 1$.

**Theorem 4.1.** *Under Assumptions 1 and 2, the conditional probabilities for system risk $S$ from (2.2) satisfy for claim $V_j$, $j \in \underline{d}$:*

(i) *for $s > v > 0$:*

$$P(S > s \mid V_j > v) = P(S > s - v) = \sum_{k=1}^{d} \tilde{\pi}_k \exp(-\tilde{\lambda}_k(s-v));$$

(ii) *asymptotically for $x \to \infty$ and $\alpha > 1$:*

$$P(S > \alpha x \mid V_j > x) \sim \tilde{\pi}_{\tilde{m}} \exp(-(\alpha-1)\tilde{\ell}x) \propto P(V_{\tilde{m}} > (\alpha-1)x),$$

*with $\tilde{m}$ and $\tilde{\ell}$ from (2.7).*

The proof is provided in Section 7.

**Remark 4.2.** Our results in Theorem 4.1(i) reveal the following interesting features: The conditional distribution for the system risk $S$ given a claim $V_j$

(i) is equal to the shifted *unconditional* distribution of the system's loss $S$,

(ii) is independent of the distinct object $j$,

(iii) depends *only* on the difference $s - v$ of the threshold values.

In particular, (i) and (iii) display a certain "no-memory" property for the GEM distribution class. These results also contribute to statistical theory, because such conditional probabilities have not been investigated for GEM distributions yet. ◇

Analogously to Theorem 4.1 we obtain the complementary result:

**Proposition 4.3.** *Under Assumptions 1 and 2 the conditional probabilities for claim $V_j$, $j \in \underline{d}$ satisfy:*

(i) *for $s > v > 0$:*

$$P(V_j > v \mid S > s) = \frac{P(V_j > v)}{P(S > s \mid S > s - v)}$$

$$= \exp(-\tilde{\lambda}_j v) \frac{\sum_{k=1}^{d} \tilde{\pi}_k \exp(-\tilde{\lambda}_k(s-v))}{\sum_{k=1}^{d} \tilde{\pi}_k \exp(-\tilde{\lambda}_k s)},$$

and for $v \geq s > 0$:
$$P(V_j > v \mid S > s) \;=\; \frac{P(V_j > v)}{P(S > s)} \;=\; \frac{\exp(-\tilde{\lambda}_j v)}{\sum_{k=1}^d \tilde{\pi}_k \exp(-\tilde{\lambda}_k s)}\,;$$

(ii) *asymptotically for $x \to \infty$ and $0 < \beta < 1$:*
$$P(V_j > \beta x \mid S > x) \;\sim\; \exp(-(\tilde{\lambda}_j - \tilde{\ell})\beta x) \;=\; \frac{P(V_j > \beta x)}{P(V_{\tilde{m}} > \beta x)}\,,$$

*and for $\beta \geq 1$:*
$$P(V_j > \beta x \mid S > x) \;\sim\; \frac{\exp(-(\beta\tilde{\lambda}_j - \tilde{\ell})x)}{\tilde{\pi}_{\tilde{m}}} \;\propto\; \frac{P(V_j > \beta x)}{P(V_{\tilde{m}} > x)}\,.$$

**Proof.** The results follow from Theorem 4.1, which implies:
$$P(V_j > v, S > s) = P(V_j > v) \cdot P(S > s - v)\,.$$
□

**Remark 4.4.** Proposition 4.3 shows that $P(V_j > v \mid S > s)$ depends (even asymptotically) on index $j$ and on the distribution of the particular claim $V_j$ in contrast to the previously investigated counterpart $P(S > s \mid V_j > v)$. ◇

Next we state the results for the conditional densities:

**Proposition 4.5.** *Under Assumptions 1 and 2 we obtain for the conditional density of the system risk $S$:*

(i) *for $s > v > 0$:*
$$f_{S \mid V_j > v}(s) = f_S(s - v)\,,$$

(ii) *asymptotically for $x \to \infty$ and $0 < \beta < 1$:*
$$f_{S \mid V_j > \beta x}(x) \;\sim\; \tilde{\pi}_{\tilde{m}} f_{V_{\tilde{m}}}((1-\beta)x) = \tilde{\pi}_{\tilde{m}} \tilde{\ell} \exp(-(1-\beta)\tilde{\ell}x)\,;$$

*and for the conditional density of claim $V_j$, $j \in \underline{d}$:*

(iii)
$$f_{V_j \mid S > s}(v) = \begin{cases} \dfrac{f_{V_j}(v)}{f_{S \mid S > s - v}(s)}\,, & \text{for } s > v > 0\,, \\[2mm] \dfrac{f_{V_j}(v)}{f_S(s)}\,, & \text{for } v \geq s > 0\,; \end{cases}$$

(iv) *asymptotically for $x \to \infty$:*
$$f_{V_j \mid S > \alpha x}(x) \;\sim\; \begin{cases} \dfrac{\tilde{\lambda}_j}{\tilde{\ell}\tilde{\pi}_{\tilde{m}}} \exp(-(\tilde{\lambda}_j - \alpha\tilde{\ell})x)\,, & \text{for } \alpha \in (0,1]\,, \\[2mm] \dfrac{\tilde{\lambda}_j}{\tilde{\ell}} \exp(-(\tilde{\lambda}_j - \tilde{\ell})x)\,, & \text{for } \alpha > 1\,. \end{cases}$$

**Proof.** These densities results follow from Theorem 4.1 and Proposition 4.3. □

We complete the analysis of conditional probabilities for large object claims by stating the results for individual agent's risk.

**Theorem 4.6.** *Under Assumptions 1 and 2, the exposure $U_i$ of agent $i$ and claim $V_j$ of object $j \in M_{(i)}$ have for $x > 0$ the following conditional distributions and asymptotic behavior for $x \to \infty$ (denoted by symbol "$\sim$"):*

(i) *for $\alpha > a_{i,j}$:*

$$\begin{aligned} P(U_i > \alpha x \mid V_j > x) &= P(U_i > (\alpha - a_{i,j})x) \\ &= \sum_{k \in M_{(i)}} \pi_{i,k} \exp(-(\alpha - a_{i,j})\lambda_{i,k}x) \\ &\sim \pi_{i,m(i)} \exp(-(\alpha - a_{i,j})\ell(i)\,x)\,; \end{aligned}$$

(ii) *for $0 < \beta < 1/a_{i,j}$:*

$$\begin{aligned} P(V_j > \beta x \mid U_i > x) &= \frac{P(V_j > \beta x)}{P(U_i > x \mid U_i > (1 - \beta a_{i,j})x)} \\ &= \exp(-\tilde{\lambda}_j \beta x) \frac{\sum_{k \in M_{(i)}} \pi_{i,k} \exp(-(1 - \beta a_{i,j})\lambda_{i,j}x)}{\sum_{k \in M_{(i)}} \pi_{i,k} \exp(-\lambda_{i,j}x)} \\ &\sim \exp(-(\tilde{\lambda}_j - a_{i,j}\ell(i))\beta x)\,; \end{aligned}$$

*and for $\beta \geq 1/a_{i,j}$:*

$$\begin{aligned} P(V_j > \beta x \mid U_i > x) &= \frac{P(V_j > \beta x)}{P(U_i > x)} = \frac{\exp(-\tilde{\lambda}_j \beta x)}{\sum_{k \in M_{(i)}} \pi_{i,k} \exp(-\lambda_{i,j}x)} \\ &\sim \frac{\exp(-(\beta\tilde{\lambda}_j - \ell(i))x)}{\pi_{i,m(i)}}\,; \end{aligned}$$

*and the conditional densities for $x > 0$:*

(iii) *for $0 < \beta < 1/a_{i,j}$:*

$$f_{U_i \mid V_j > \beta x}(x) = f_{U_i}((1 - \beta a_{i,j})x) \sim \pi_{i,m(i)}\ell(i) \exp(-(1 - \beta a_{i,j})x)\,;$$

(iv)

$$f_{V_j \mid U_i > \alpha x}(x) = \begin{cases} \dfrac{f_{V_j}(x)}{f_{U_i}(\alpha x)} \sim \tilde{\lambda}_j \exp(-(\tilde{\lambda}_j - \alpha\ell(i))x) & \text{for } 0 < \alpha \leq a_{i,j}\,, \\ \dfrac{f_{V_j}(x)}{f_{U_i \mid U_i > (\alpha - a_{i,j})x}(\alpha x)} \sim \tilde{\lambda}_j \exp(-(\tilde{\lambda}_j - \ell(i))x) & \text{for } \alpha > a_{i,j}\,. \end{cases}$$

**Proof.** The results follow analogously to the proofs of Theorem 4.1 and Propositions 4.3, 4.5, whereby here we apply $U_i = \sum_{k \in M_{(i)}} X_{i,k}$ with $X_{i,k} = a_{i,k} V_k$ which are stochastically independent and exponentially distributed with parameters $\lambda_{i,k}$.

□

**Remark 4.7.** We provide some economic interpretations of our findings, in particular how our results are applicable to system regulation. Theorems 4.1 and 4.6(i) point out the following characteristic properties for financial systems with exponential claims:

The marginal distribution of the system's loss $S$ (or a single agent's loss $U_i$) contains *all* information to quantify the effect of (large) claims on the system (or individual) risk. This holds not only asymptotically for large claims, but also exactly for all $s, v > 0$: $P(S > s \mid V_j > v) = P(S > s - v)$.

This "no-memory" property of the GEM distribution class leads to a substantial simplification for system regulation. Only the marginal distribution of the system (or individual) risk has to be known in order to quantify system (or individual) stability with respect to stress situations caused by large claims. Thereby, it is irrelevant which particular object causes troubles.

The other way around, by evaluating the impact of a high system's loss $S$ on the distribution of a single claim $V_j$, we obtain that it depends specifically on the claim's stochastic properties and characteristic parameters. Nevertheless, in this case the conditional distributions can be expressed via marginal ones for both $S$ and $V_j$, as Proposition 4.3 implies:

$$P(V_j > v \mid S > s) = \frac{P(V_j > v) P(S > s - v)}{P(S > s)}, \quad s, v > 0.$$

The qualitative difference between $P(S > s \mid V_j > v)$ and $P(V_j > v \mid S > s)$ as well as between $P(U_i > u \mid V_j > v)$ and $P(V_j > v \mid U_i > u)$ is based on the following intuition: events $\{S > s\}$ or $\{U_i > u\}$ do not specify which object causes which loss. Such events may comprise very different scenarios for the vector of object claims $(V_1, \ldots, V_d)$, e.g. scenarios with one or more *large* claims as well as those where none of the object claims is large but $S$ or $U_i$ exceeds a high threshold by a cumulation effect. ◇

## 5 Mutual influence of individual and system risks with respect to portfolio diversification

In this section we deduce the conditional distribution for the system risk in case that some agent confronts a stress situation and, conversely, the conditional distribution of an agent's exposure in situations of large system's losses.

By analogy to Theorem 4.1 and Proposition 4.3, replacing the single object claims $V_j$ by generically structured portfolios $U_i$ as in (2.1) requires to compute numerous distinct cases. Hence, we focus on *homogeneous portfolios* as it is common in the literature (see e.g. Brechmann et al. [4], Geluk et al. [8], Ibragimov [10]), where the considered agent $i$ selects arbitrary objects for his portfolio but holds them equally weighted, i.e.,

$$a_{i,j} = a \quad \text{for all } j \in M_{(i)} \text{ and some } 0 < a \leq 1\,. \tag{5.1}$$

We illustrate the technical results in Theorem 5.1 and Proposition 5.2 by the two extreme portfolios (cf. Section 5.2): the concentrated portfolio (A) with $|M_{(i)}| = 1$ and the totally diversified portfolio (B) with $|M_{(i)}| = d$ in Corollaries 5.3 to 5.6. These allow us to quantify diversification effects in Remark 5.7 and to formulate in Theorem 5.8 a criterion to decide whether portfolio diversification is beneficial for individual agents with respect to a systemic crisis.

We extend our notation characterizing the objects which agent $i$ selects or not. The set $\underline{d}$ of all object indices is divided into the set $M_{(i)}$ from (2.4) of objects selected by agent $i$ and its complement

$$M_{(\neg i)} := \underline{d} \setminus M_{(i)}\,, \tag{5.2}$$

which contains the objects not selected by agent $i$. Moreover, we define the sets

$$M_{(i \setminus j)} := M_{(i)} \setminus \{j\} \quad \text{and} \quad M_{(\neg i \cup j)} := M_{(\neg i)} \cup \{j\} = \underline{d} \setminus M_{(i \setminus j)}.$$

Analogously to $\tilde{\ell}$ from (2.7) we define

$$\tilde{\ell}(i) := \tilde{\lambda}_{\tilde{m}(i)} := \min_{j \in M_{(i)}} \tilde{\lambda}_j \quad \text{and} \quad \tilde{\ell}(\neg i) := \tilde{\lambda}_{\tilde{m}(\neg i)} := \min_{j \in M_{(\neg i)}} \tilde{\lambda}_j\,, \tag{5.3}$$

with $\tilde{\ell}(\neg i) := \infty$ if $M_{(\neg i)}$ is empty. Extending the notion of $\tilde{\pi}_j$ from (3.5) to subsets of $\underline{d}$, we denote the corresponding mixing proportions for $j \in M_{(*)}$ and $* \in \{i, \neg i, \neg i \cup j\}$ by

$$\tilde{\pi}_{*,j} := \begin{cases} \displaystyle\prod_{k \in M_{(*)} \setminus \{j\}} \frac{\tilde{\lambda}_k}{\tilde{\lambda}_k - \tilde{\lambda}_j} & \text{for } |M_{(*)}| > 1\,, \\ 1 & \text{otherwise}\,. \end{cases} \tag{5.4}$$

These mixing proportions satisfy

$$\tilde{\pi}_{i,j} \cdot \tilde{\pi}_{\neg i \cup j, j} = \tilde{\pi}_j \quad \text{for all } i \in \underline{n}, j \in M_{(i)}\,. \tag{5.5}$$

## 5.1 Conditional results on individual and system risks

In the following theorem we state the conditional distribution of the system's loss $S$ given that some agent $i$ suffices a loss of at least $u$; the proof is provided in Section 7.

**Theorem 5.1.** *Let Assumptions 1, 2 and (5.1) hold. Then*

(i) *for $s > u/a > 0$:*

$$P(S > s \mid U_i > u) = \frac{\sum_{j \in M_{(i)}} \tilde{\pi}_{i,j} P(V_j > u/a) P(\sum_{k \in M_{(\neg i \cup j)}} V_k > s - u/a)}{\sum_{j \in M_{(i)}} \tilde{\pi}_{i,j} P(V_j > u/a)}$$

$$= \frac{\sum_{j \in M_{(i)}} \sum_{k \in M_{(\neg i \cup j)}} \tilde{\pi}_{i,j} \tilde{\pi}_{\neg i \cup j, k} \exp(-\frac{\tilde{\lambda}_j - \tilde{\lambda}_k}{a} u - \tilde{\lambda}_k s)}{\sum_{j \in M_{(i)}} \tilde{\pi}_{i,j} \exp(-\frac{\tilde{\lambda}_j}{a} u)};$$

(ii) *asymptotically for $x \to \infty$ and $\alpha > 1/a$:*

$$P(S > \alpha x \mid U_i > x) \sim \tilde{\pi}_{\neg i \cup \tilde{m}(i), \tilde{m}} \exp(-(\alpha - 1/a)\tilde{\ell} x).$$

Conversely, we obtain:

**Proposition 5.2.** *Let Assumptions 1, 2 and (5.1) hold. Then*

(i) *for $s > u/a > 0$:*

$$P(U_i > u \mid S > s) = \frac{\sum_{j \in M_{(i)}} \tilde{\pi}_{i,j} P\left(V_j > \frac{u}{a}\right) P\left(\sum_{k \in M_{(\neg i \cup j)}} V_k > s - \frac{u}{a}\right)}{\sum_{q \in \underline{d}} \tilde{\pi}_q P(V_q > s)}$$

$$= \frac{\sum_{j \in M_{(i)}} \sum_{k \in M_{(\neg i \cup j)}} \tilde{\pi}_{i,j} \tilde{\pi}_{\neg i \cup j, k} \exp\left(-\frac{\tilde{\lambda}_j - \tilde{\lambda}_k}{a} u - \tilde{\lambda}_k s\right)}{\sum_{q \in \underline{d}} \tilde{\pi}_q \exp(-\tilde{\lambda}_q s)},$$

*and for $u \geq a s > 0$:*

$$P(U_i > u \mid S > s) = \frac{\sum_{j \in M_{(i)}} \tilde{\pi}_{i,j} P\left(V_j > \frac{u}{a}\right)}{\sum_{q \in \underline{d}} \tilde{\pi}_q P(V_q > s)} = \frac{\sum_{j \in M_{(i)}} \tilde{\pi}_{i,j} \exp\left(-\frac{\tilde{\lambda}_j}{a} u\right)}{\sum_{q \in \underline{d}} \tilde{\pi}_q \exp(-\tilde{\lambda}_q s)};$$

(ii) *asymptotically for $x \to \infty$ and $0 < \beta < a$:*

- *if $\tilde{\ell}(i) < \tilde{\ell}(\neg i)$:*

$$P(U_i > \beta x \mid S > x) \to 1,$$

- *if $\tilde{\ell}(i) > \tilde{\ell}(\neg i)$:*

$$P(U_i > \beta x \mid S > x) \sim \prod_{j \in M_{(i \setminus \tilde{m}(i))}} \frac{\tilde{\lambda}_j - \tilde{\ell}(\neg i)}{\tilde{\lambda}_j - \tilde{\ell}(i)} \exp\left(-(\tilde{\ell}(i) - \tilde{\ell}(\neg i)) \frac{\beta}{a} x\right);$$

*and for $\beta \geq a$:*

$$P(U_i > \beta x \mid S > x) \sim \frac{\tilde{\pi}_{i, \tilde{m}(i)}}{\tilde{\pi}_{\tilde{m}(i)}} \exp\left(-\left(\frac{\beta}{a} \tilde{\ell}(i) - \tilde{\ell}\right) x\right).$$

The different cases for the asymptotic results in Proposition 5.2(ii) arise depending on whether the dominant claim of the system (according to Theorem 3.5) with the smallest parameter $\tilde{\ell}$ is a part of the agent's portfolio.

**Proof.** The results in Proposition 5.2 follow from Theorem 5.1, whose proof (cf. (7.4)) gives:

$$P(U_i > u, S > s)$$
$$= \begin{cases} \sum_{j \in M_{(i)}} \tilde{\pi}_{i,j} P(V_j > u/a) \, P\Big(\sum_{k \in M_{(\neg i \cup j)}} V_k > s - u/a\Big) & \text{for } s > u/a, \\ \sum_{j \in M_{(i)}} \tilde{\pi}_{i,j} P(V_j > u/a) & \text{for } u \geq a s. \end{cases}$$

□

In Theorem 5.1 and Proposition 5.2 we quantify individual and system risks by their conditional loss distributions, which allow us to measure interdependencies within the system. These results display the impact of different types of stress situations such as a distressed agent or a systemic crisis and, hence, are useful for stress testing of financial institutions (see e.g. Brechmann et al. [4]).

## 5.2 Analysis of diversification benefits

Portfolio selection and diversification refer to central concepts in financial risk management. We provide an assessment of possible diversification benefits by considering two important extreme cases, cf. Ibragimov [10]:

**(A) Concentrated portfolio:** agent $i$ selects only object $j$;
i.e., $a_{i,j} = a$ for some $0 < a \leq 1$, $a_{i,k} = 0$ for all $k \in \underline{d} \setminus \{j\}$.

**(B) Totally diversified portfolio:** agent $i$ selects *all* objects and with the *same* weights;
i.e., $a_{i,k} = a$ for all $k \in \underline{d}$ and some $0 < a \leq 1$.

For these portfolios (A) and (B) we obtain the following results in Corollaries 5.3 – 5.6 as special cases of Theorem 5.1 and Proposition 5.2.

We start with the concentrated portfolio (A) with $U_i = a_{i,j} V_j$ which implies $M_{(i)} = \{j\}$ and, hence, $\tilde{\ell}(i) = \tilde{\lambda}_j$, $\tilde{\pi}_{i,j} = 1$ and $\tilde{\pi}_{\neg i \cup j, k} = \tilde{\pi}_k$ for all $k \in \underline{d}$. Then we apply property (5.5) for the mixing proportions and obtain:

**Corollary 5.3** (Concentrated portfolio). *Under Assumptions 1, 2, for portfolio (A) we find:*

*(i) for $s > u/a > 0$:*

$$P(S > s \mid U_i > u) = P\left(S > s - \frac{u}{a}\right) = \sum_{k=1}^{d} \tilde{\pi}_k \exp\left(-\tilde{\lambda}_k \left(s - \frac{u}{a}\right)\right);$$

(ii) *asymptotically for $x \to \infty$ and $\alpha > 1/a$:*

$$P(S > \alpha x \mid U_i > x) \sim \tilde{\pi}_{\tilde{m}} \exp\left(-\left(\alpha - \frac{1}{a}\right)\tilde{\ell} x\right).$$

And complementary:

**Corollary 5.4** (Concentrated portfolio). *Under Assumptions 1, 2, for portfolio (A) we find:*

(i) *for $s > u/a > 0$:*

$$P(U_i > u \mid S > s) = \frac{P(U_i > u)}{P\left(S > s \mid S > s - \frac{u}{a}\right)}$$
$$= \frac{\sum_{k=1}^{d} \tilde{\pi}_k \exp\left(-\frac{\tilde{\lambda}_j - \tilde{\lambda}_k}{a} u - \tilde{\lambda}_k s\right)}{\sum_{k=1}^{d} \tilde{\pi}_k \exp(-\tilde{\lambda}_k s)},$$

*and for $u \geq a\, s > 0$:*

$$P(U_i > u \mid S > s) = \frac{P(U_i > u)}{P(S > s)} = \frac{\exp\left(-\frac{\tilde{\lambda}_j}{a} u\right)}{\sum_{k=1}^{d} \tilde{\pi}_k \exp(-\tilde{\lambda}_k s)};$$

(ii) *asymptotically for $x \to \infty$ and $0 < \beta < a$:*

- *if $\tilde{\lambda}_j = \min_{k \in \underline{d}} \tilde{\lambda}_k$:*

$$P(U_i > \beta x \mid S > x) \to 1,$$

- *if $\tilde{\lambda}_j \neq \min_{k \in \underline{d}} \tilde{\lambda}_k$:*

$$P(U_i > \beta x \mid S > x) \sim \exp\left(-(\tilde{\lambda}_j - \tilde{\ell})\frac{\beta}{a} x\right);$$

*and for $\beta \geq a$:*

$$P(U_i > \beta x \mid S > x) \sim \frac{1}{\tilde{\pi}_j} \exp\left(-\left(\frac{\beta}{a}\tilde{\lambda}_j - \tilde{\ell}\right) x\right).$$

Next, we analyze the totally diversified portfolio (B). We obtain the following results from Theorem 5.1 and Proposition 5.2 with $U_i = a \sum_{j=1}^{d} V_j = a\, S$ which implies $M_{(i)} = \underline{d}$, $\tilde{\ell}(i) = \tilde{\ell}$, $\tilde{\pi}_{i,j} = \tilde{\pi}_j$ for all $j \in \underline{d}$ and $\tilde{\pi}_{\neg i \cup j, k} = 1$. The set $M_{(\neg i)}$ (of non-selected objects) is empty for portfolio (B), so we have $\tilde{\ell}(\neg i) = \infty$, cf. (5.3).

**Corollary 5.5** (Totally diversified portfolio). *Under Assumptions 1, 2, for portfolio (B) we find:*

(i) *for $s > u/a > 0$:*

$$P(S > s \mid U_i > u) = P\left(S > s \mid S > \frac{u}{a}\right) = \frac{\sum_{j=1}^{d} \tilde{\pi}_j \exp(-\tilde{\lambda}_j s)}{\sum_{j=1}^{d} \tilde{\pi}_j \exp\left(-\frac{\tilde{\lambda}_j}{a} u\right)};$$

(ii) *asymptotically for $x \to \infty$ and $\alpha > 1/a$:*

$$P(S > \alpha x \mid U_i > x) \sim \exp\left(-\left(\alpha - \frac{1}{a}\right)\tilde{\ell} x\right).$$

And complementary:

**Corollary 5.6** (Totally diversified portfolio). *Under Assumptions 1, 2, for portfolio (B) we find:*

(i) *for $u \geq a\, s > 0$:*

$$P(U_i > u \mid S > s) = P\left(S > \frac{u}{a} \mid S > s\right) = \frac{\sum_{j=1}^{d} \tilde{\pi}_j \exp\left(-\frac{\tilde{\lambda}_j}{a} u\right)}{\sum_{j=1}^{d} \tilde{\pi}_j \exp(-\tilde{\lambda}_j s)};$$

(ii) *asymptotically for $x \to \infty$ and $\beta \geq a$ it holds:*

$$P(U_i > \beta x \mid S > x) \sim \exp\left(-\left(\frac{\beta}{a} - 1\right)\tilde{\ell} x\right).$$

Our results in Corollaries 5.3 – 5.6 for the two extreme cases of portfolio selection give interesting insights for diversification policy which are discussed in the following Remark 5.7 and in Theorem 5.8.

**Remark 5.7.** The impact of an agent's large exposure on the system risk is of interest for system regulation. This issue has been considered in Theorem 5.1 and Corollaries 5.3, 5.5, which show that for a homogeneous portfolio of agent $i \in \underline{n}$ with weights $a_{i,j} \in \{0, a\}$ for $0 < a \leq 1$ the conditional probabilities for the system's loss are all asymptotically proportional:

$$P(S > \alpha x \mid U_i > x) \sim C(M_{(i)}) \exp(-(\alpha - 1/a)\tilde{\ell} x) \quad \text{for} \quad x \to \infty,$$

with $C(M_{(i)}) := \tilde{\pi}_{\neg i \cup \tilde{m}(i), \tilde{m}}$. We find that $C(M_{(i)}) > C(M_{(k)})$ for $M_{(i)} \subset M_{(k)}$. The proportionality constant $C(M_{(i)})$ is equal to 1 if agent $i$ holds a totally diversified portfolio (B) including all available objects, and increases in the number of objects not selected by agent $i$ up to the value $\tilde{\pi}_{\tilde{m}}$ for the concentrated (one-object) portfolio (A).

Thus, given that a single agent's exposure exceeds some high threshold, the probability for a large system's loss is higher for a smaller portfolio's diversification degree of the agent. This holds because, if the sum of several shared objects $U_i = a \sum_{j \in M_{(i)}} V_j$ exceeds some high threshold, this does not require that we have large claims on single objects, as it can be a cumulation effect. However, if a single $aV_j$ exceeds a high threshold, it is a clear cause of a system's loss. ◇

Next we show that for systems with exponential claims the typical recommendation for portfolio selection has to be modified: Diversification is *not always beneficial* in view of systemic crisis situations. The following theorem gives a criterion for agents to decide whether concentration or diversification is beneficial.

**Theorem 5.8.** *Let agent 1 hold a concentrated portfolio (A) by selecting a single object $j$ with weight $a$ and agent 2 hold a totally diversified portfolio (B) by selecting all $d$ objects with weights $a/d$, respectively. In order to minimize the individual risk with respect to situations of large system's losses, i.e. to minimize the conditional probability of*
$$P_i(x) := P(U_i > \beta x \mid S > x), \quad \beta > 0, \ i = 1, 2$$
*as $x \to \infty$, we deduce the following decision criterion:*
*Portfolio concentration (agent 1) is beneficial in contrast to diversification (agent 2) if and only if the ratio of parameters satisfies:*

$$\frac{\tilde{\lambda}_j}{\tilde{\ell}} > Q_{d,\beta}, \quad \text{with } Q_{d,\beta} := \begin{cases} 1 & \text{for } 0 < \beta < a/d, \\ 1 + d - a/\beta & \text{for } a/d \leq \beta < a, \\ d & \text{for } \beta \geq a, \end{cases} \quad (5.6)$$

*where $\tilde{\lambda}_j$ is the parameter of object $j$ and $\tilde{\ell} = \min_{k \in \underline{d}} \tilde{\lambda}_k$.*

**Proof.** Corollaries 5.4(ii) with weight $a$ and 5.6(ii) with weight $a/d$ imply that:

$$P_1(x) \sim \begin{cases} \exp(-(\tilde{\lambda}_j - \tilde{\ell})\beta x/a) & \text{for } 0 < \beta < a \\ \exp(-(\beta\tilde{\lambda}_j/a - \tilde{\ell})x)/\tilde{\pi}_j & \text{for } \beta \geq a \end{cases}$$

$$P_2(x) \sim \begin{cases} 1 & \text{for } 0 < \beta < a/d \\ \exp(-(\beta d/a - 1)\tilde{\ell}x) & \text{for } \beta \geq a/d \end{cases}$$

Consequently, we obtain:

– for $0 < \beta < a/d$:
  $P_1(x) = o(P_2(x))$ if $\tilde{\lambda}_j \neq \tilde{\ell}$ (i.e., $\tilde{\lambda}_j > \tilde{\ell}$) and $P_1(x) \sim P_2(x)$ if $\tilde{\lambda}_j = \tilde{\ell}$;

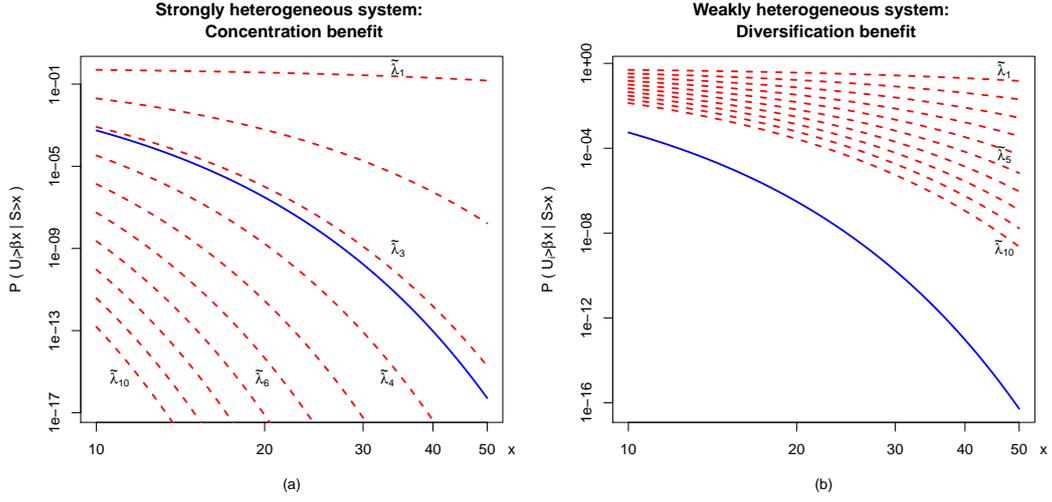

**Figure 5.1:** Log-log-plot: comparison of conditional probabilities for ten concentrated portfolios for objects $j \in \underline{d} = \{1, \ldots, 10\}$ (dashed lines) and for totally diversified portfolio (solid line) with $\beta = 0.4$, $a = 0.25$: plot (a) for scenario (i) with strongly heterogeneous parameters $\tilde{\lambda}_j$ and plot (b) for scenario (ii) with weakly heterogeneous parameters $\tilde{\lambda}_j$; cf. Example 5.9.

– for $a/d \leq \beta < a$:

$$P_1(x) = o(P_2(x)) \quad \Leftrightarrow \quad -\frac{\tilde{\lambda}_j \beta}{a} + \frac{\tilde{\ell}\beta}{a} + \frac{\beta d \tilde{\ell}}{a} - \tilde{\ell} < 0 \quad \Leftrightarrow \quad \frac{\tilde{\lambda}_j}{\tilde{\ell}} > 1 + d - \frac{a}{\beta};$$

– for $\beta \geq a$:

$$P_1(x) = o(P_2(x)) \quad \Leftrightarrow \quad -\frac{\tilde{\lambda}_j \beta}{a} + \frac{\tilde{\ell}\beta d}{a} < 0 \quad \Leftrightarrow \quad \frac{\tilde{\lambda}_j}{\tilde{\ell}} > d.$$

□

As a consequence of Theorem 5.8, portfolio diversification is usually preferable for agents if objects refer to similar risk classes. However, in systems with strongly heterogeneous object claims (very different risk classes, e.g. governmental bonds and derivative products), concentrating on few objects identified by criterion (5.6) can be advantageous for agents in situations of large system's losses. We demonstrate this effect in the following example and the corresponding Figure 5.1.

**Example 5.9.** For a system of $d = 10$ objects we compare the conditional survival functions $P_1$ for the ten portfolios concentrated each on a single object $j \in \underline{d}$ and $P_2$ for the totally diversified portfolio of all ten objects for two different scenarios: (i) system with strongly heterogeneous objects, (ii) system with weakly heterogeneous objects. In scenario (i) the claims have parameters $(\tilde{\lambda}_1, \tilde{\lambda}_2, \ldots, \tilde{\lambda}_{10}) = (0.05, 0.25, \ldots 1.85)$ with step 0.2, in scenario (ii) they have parameters $(\tilde{\lambda}_1, \tilde{\lambda}_2, \ldots,$ $\tilde{\lambda}_{10}) = (0.05, 0.75, \ldots 0.275)$ with step $0.025$. These scenarios are suitable for comparison as the risk-dominant parameter is $\tilde{\ell} = 0.05$ equally for (i) and (ii).

In Figure 5.1 we plot for $\beta = 0.4$, $a = 0.25$ the survival functions of type $P_1$ for ten concentrated portfolios (dashed lines) and those of type $P_2$ for the totally diversified portfolio (solid line). It illustrates how our criterion in Theorem 5.8 applies: In scenario (ii) the parameter ratio is $\tilde{\lambda}_j/\tilde{\ell} < d$ for all $j \in \underline{d}$, and the criterion gives that diversification is most beneficial here. In contrast, in scenario (i) it holds $\tilde{\lambda}_j/\tilde{\ell} > d$ if and only if $j \geq 4$, which implies that not diversification but concentration on objects $k \in \{4, 5, \ldots, 10\}$ is the favorable strategy here for agents to minimize their individual risks in case of systemic crisis situations. $\diamondsuit$

## 6 Summary and comparison of results for light and heavy tails

In this section, we contrast our results for a system with light-tailed claims $V_j \in \text{EXP}(\lambda_j)$ for $j \in \underline{d}$ to the earlier results under the assumption of heavy-tailed claims. The papers of Kley et al. [15, 16] are very appropriate for a comparison as they investigate systems of the same structure. The only difference to our framework is that the object claims are assumed to be heavy-tailed with asymptotically proportional tail decay, in particular they have Pareto tails:

$$P(V_j > x) \propto x^{-\gamma} \quad \text{for } x \to \infty, \tag{6.1}$$

with same parameter $\gamma > 0$ for all $j \in \underline{d}$; written as $V_j \in \text{PAR}(\gamma)$.

Our comparison addresses three issues: (I) scaling and aggregation of claims, (II) individual and system's risks, (III) conditional distributions and diversification effects.

**(I)** We start by contrasting the qualitatively different behavior for exponential $\text{EXP}(\lambda)$ and for Pareto $\text{PAR}(\gamma)$ claims under scaling and aggregation. In particular we underscore the different role of parameters $\lambda$ and $\gamma$ which both describe the tail decay of the claims' survival functions but act differently as scale and shape parameters, respectively:

– Scaling with $a > 0$: for exponential claims the tail decay changes, for Pareto claims it remains unchanged:

$$\begin{aligned} V \in \text{EXP}(\lambda) &\Rightarrow aV \in \text{EXP}(\lambda/a), \\ V \in \text{PAR}(\gamma) &\Rightarrow aV \in \text{PAR}(\gamma). \end{aligned}$$

– Aggregation: summing up exponential claims leads to classes of GEM or generalized Erlang mixture distributions, for Pareto claims the distribution type PAR remains with unchanged $\gamma$:

$$V_i \in \text{EXP}(\lambda_i), i = 1, 2 \quad \Rightarrow \quad V_1 + V_2 \in \text{GEM}(\lambda_1, \lambda_2),$$
$$V_1, V_2 \in \text{PAR}(\gamma) \quad \Rightarrow \quad V_1 + V_2 \in \text{PAR}(\gamma).$$

These two points also imply that in the EXP-framework we cannot restrict our analysis to object claims with the same tail decay, as in the PAR-framework in Kley et al. [15, 16]. They assumed PAR($\gamma$) with same parameter $\gamma$ for all object claims because claims with larger parameter are asymptotically negligible. We have shown that this is not the case for EXP-framework. Theorem 3.5 proves that object claims which are asymptotically negligible for the system might be dominant ones for some agents' exposures. Consequently, the framework of Jiang and Tang [13] where the claims are assumed to be independent exponentially distributed with the *same* parameter $\lambda$ and the aggregated risk is analyzed using the Erlang distribution cannot be applied in our context; i.e., to systems of agents sharing claims.

**(II)** Next, we compare individual and system risks for both EXP-framework with distinct parameters according to Assumption 2 and PAR-framework:

For exponential claims, we prove that the individual risk of an agent is determined asymptotically by the dominant claim in his portfolio. The survival functions of the agents' exposures have generically different tail decays, and can differ from those of the system's loss. This means that some individual risks are asymptotically negligible compared to others and, in particular, to the system risk. Generically it holds

$$P(U_i > x) = o(P(S > x)) \quad \text{for} \quad x \to \infty,$$

as we have shown in Theorem 3.5, cf. also Remark 3.6.

In contrast, for PAR-framework individual and system risks are asymptotically proportional (Th. 3.2 in Kley et al. [15]):

$$P(U_i > x) \sim C_i \, x^{-\gamma}, \qquad P(S > x) \sim C_S \, x^{-\gamma}, \tag{6.2}$$

with constant $C_i > 0$ depending on *all* objects selected by agent $i$ and constant $C_S > 0$ depending on *all* objects in the network. In contrast, for EXP-framework only single objects determine the asymptotic tails of individual and system risks; cf. Theorem 3.5.

**(III)** The impact of single object claims as in Section 4 has not been investigated in Kley et al. [15, 16]; however, the "no-memory" property of the EXP-framework for the conditional distribution of $S$ (in the GEM distribution class) given some claim $V_j$ shown in Theorem 4.1 and Remark 4.2 does not hold for PAR. In contrast to our results in Section 5 for the EXP-framework, Kley et al. [16, Th. 2.4, Cor. 2.5] obtain for the PAR-framework with homogeneous portfolios defined in (5.1) the following statements on the mutual dependence of individual and system risk for $x \to \infty$:

$$P(S > \alpha x \mid U_i > x) \;\to\; (\alpha a)^{-\gamma} \quad \text{for } \alpha > 1/a\,, \tag{6.3}$$

$$P(U_i > \beta x \mid S > x) \;\to\; \begin{cases} 1 & \text{for } \beta < a \\ (a/\beta)^\gamma & \text{for } \beta \geq a \end{cases} \tag{6.4}$$

Hence, for PAR-claims the conditional probabilities of individual and system risks converge in all cases to positive limits. This shows a qualitative difference to EXP-claims, where we have in several cases the limit value zero (see Theorem 5.1 and Proposition 5.2). Moreover in the PAR-framework, neither the specific portfolio structure of agent $i$ nor the number of selected objects influences the limits in (6.3) and (6.4). Consequently, the degree of portfolio diversification does not affect these conditional distributions in the PAR-framework. This is different in the EXP-framework, as we obtained in Corollaries 5.3 – 5.6 significant diversification effects. These findings allow for interesting interpretations given in Remark 5.7 and enable the formulation of a criterion in Theorem 5.8 to decide whether diversification brings advantages to a single agent holding exponential claims in case of a systemic crisis.

## 7 Proofs of Theorems 4.1 and 5.1

**Proof of Theorem 4.1.** For the joint probability we compute for $\alpha > 1$, $x > 0$:

$$\begin{aligned}
&P(S > \alpha x, V_j > x) \\
&= P\Big(\sum_{k=1}^d V_k > \alpha x, V_j > x\Big) = \int_x^\infty P\Big(\sum_{\substack{k \in \underline{d} \\ k \neq j}} V_k > \alpha x - u \mid V_j = u\Big) f_{V_j}(u)\,\mathrm{d}u \\
&= \int_x^{\alpha x} P\Big(\sum_{\substack{k \in \underline{d} \\ k \neq j}} V_k > \alpha x - u\Big) f_{V_j}(u)\,\mathrm{d}u + \int_{\alpha x}^\infty f_{V_j}(u)\,\mathrm{d}u \;=:\; I_1 + I_2
\end{aligned} \tag{7.1}$$

as the claims $V_k$ are stochastically independent. The integral $I_2$ is equal to $\exp(-\tilde{\lambda}_j \alpha x)$. For the integral $I_1$ we apply that $\sum_{k \neq j} V_k$ follows a GEM distribution (analogously to $S$); however, its mixing proportions do not coincide with the $\tilde{\pi}_k$ from $S = \sum_{k \in \underline{d}} V_k$, instead its mixing proportions are as follows:

$$\prod_{\substack{l \in \underline{d} \\ l \neq j, l \neq k}} \frac{\tilde{\lambda}_l}{\tilde{\lambda}_l - \tilde{\lambda}_k}, \quad k \in \underline{d} \setminus \{j\}\,. \tag{7.2}$$

Hence, dropping the summand $V_j$ does not only reduce the number of mixing proportions from $\underline{d}$ to $\underline{d}-1$, but also changes each of the mixing proportions. We derive:

$$
\begin{aligned}
I_1 &= \int_x^{\alpha x} P\Big(\sum_{\substack{k\in\underline{d}\\k\neq j}} V_k > \alpha x - u\Big) f_{V_j}(u)\mathrm{d}u \\
&= \int_x^{\alpha x} \sum_{\substack{k\in\underline{d}\\k\neq j}} \prod_{\substack{l\in\underline{d}\\l\neq j, l\neq k}} \frac{\tilde{\lambda}_l}{\tilde{\lambda}_l - \tilde{\lambda}_k} \exp(-\tilde{\lambda}_k(\alpha x - u))\tilde{\lambda}_j \exp(-\tilde{\lambda}_j u)\mathrm{d}u \\
&= \sum_{\substack{k\in\underline{d}\\k\neq j}} \tilde{\lambda}_j \prod_{\substack{l\in\underline{d}\\l\neq j, l\neq k}} \frac{\tilde{\lambda}_l}{\tilde{\lambda}_l - \tilde{\lambda}_k} \exp(-\tilde{\lambda}_k \alpha x) \int_x^{\alpha x} \exp((\tilde{\lambda}_k - \tilde{\lambda}_j)u)\mathrm{d}u \\
&= \sum_{\substack{k\in\underline{d}\\k\neq j}} \frac{\tilde{\lambda}_j}{\tilde{\lambda}_k - \tilde{\lambda}_j} \frac{\prod_{\substack{l\in\underline{d}\\l\neq j, l\neq k}} \tilde{\lambda}_l}{\prod_{\substack{l\in\underline{d}\\l\neq j, l\neq k}} (\tilde{\lambda}_l - \tilde{\lambda}_k)} \exp(-\tilde{\lambda}_k \alpha x) \\
&\qquad\qquad\qquad\qquad\qquad\qquad \cdot \Big[\exp((\tilde{\lambda}_k - \tilde{\lambda}_j)\alpha x) - \exp((\tilde{\lambda}_k - \tilde{\lambda}_j)x)\Big] \\
&= \sum_{\substack{k\in\underline{d}\\k\neq j}} \frac{\tilde{\lambda}_j \prod_{\substack{l\in\underline{d}\\l\neq j, l\neq k}} \tilde{\lambda}_l}{(\tilde{\lambda}_k - \tilde{\lambda}_j)\prod_{\substack{l\in\underline{d}\\l\neq j, l\neq k}} (\tilde{\lambda}_l - \tilde{\lambda}_k)} \Big[\exp(-\tilde{\lambda}_j \alpha x) \\
&\qquad\qquad\qquad\qquad\qquad\qquad\qquad - \exp(-(\tilde{\lambda}_j + \tilde{\lambda}_k(\alpha-1))x)\Big] \\
&= \exp(-\tilde{\lambda}_j x)\sum_{\substack{k\in\underline{d}\\k\neq j}} \tilde{\pi}_k \Big[\exp(-\tilde{\lambda}_k(\alpha-1)x) - \exp(-\tilde{\lambda}_j(\alpha-1)x)\Big],
\end{aligned}
$$

where we have used definition (3.5) in the last step. Together with (7.1) this yields

$$
\begin{aligned}
P(S > \alpha x \mid V_j > x) &= \sum_{\substack{k\in\underline{d}\\k\neq j}} \tilde{\pi}_k \exp(-\tilde{\lambda}_k(\alpha-1)x) \\
&\qquad\qquad + \Big(1 - \sum_{\substack{k\in\underline{d}\\k\neq j}} \tilde{\pi}_k\Big) \exp(-\tilde{\lambda}_j(\alpha-1)x) \\
&= \sum_{k\in\underline{d}} \tilde{\pi}_k \exp(-\tilde{\lambda}_k(\alpha-1)x) = P(S > (\alpha-1)x),
\end{aligned}
$$

where the property $\sum_{k\in\underline{d}}\tilde{\pi}_k = 1$ from (3.3) for the mixing proportions is applied. Asymptotically for $x\to\infty$, it holds:

$$
P(S > \alpha x \mid V_j > x) \sim \tilde{\pi}_{\tilde{m}}\exp(-(\alpha-1)\tilde{\ell}x) = \tilde{\pi}_{\tilde{m}}P(V_{\tilde{m}} > (\alpha-1)x).
$$

$\square$

**Proof of Theorem 5.1.** One could write the joint probability of $S$ and $U_i$ by extending the method applied in the proof of Theorem 4.1, where in (7.1) the joint probability of $S$ and $V_j$ is calculated by integration of conditional probabilities given the claim value using the stochastic independence of $V_j$ from the other object claims. However, this would lead to the following non-trivial $|M_{(i)}|$-dimensional integrals:

$$P\Big(S > \alpha x, U_i > x\Big) = P(\sum_{j \in \underline{d}} V_j > \alpha x, \sum_{j \in M_{(i)}} V_j > x/a)$$

$$= \int \cdots \int_{\substack{v_j \geq 0, j \in M_{(i)} \\ \text{with } \sum v_j > x/a}} P(\sum_{j \in \underline{d}} V_j > \alpha x \mid V_j = v_j, j \in M_{(i)}) \cdot \prod_{j \in M_{(i)}} (f_{V_j}(v_j) \, dv_j)$$

$$= \int \cdots \int_{\substack{v_j \geq 0, j \in M_{(i)} \\ \text{with } x/a < \sum v_j < \alpha x}} P(\sum_{k \in M_{(\neg i)}} V_k > \alpha x - \sum_{j \in M_{(i)}} v_j) \cdot \prod_{j \in M_{(i)}} (f_{V_j}(v_j) \, dv_j)$$

$$+ \int \cdots \int_{\substack{v_j \geq 0, j \in M_{(i)} \\ \text{with } \sum v_j > \alpha x}} \prod_{j \in M_{(i)}} (f_{V_j}(v_j) \, dv_j).$$

To avoid this multi-dimensional integration over all combinations for the elements of $(v_j)_{j \in M_{(i)}} \in [0, \infty)^{|M_{(i)}|}$ where their sum $\sum_j (v_j)$ lays in some distinct interval, we propose the following idea. We introduce the random variables

$$W_i := U_i/a = \sum_{j \in M_{(i)}} V_j \quad \text{and} \quad \overline{W}_i := S - W_i = \sum_{k \in M_{(\neg i)}} V_j.$$

Then we apply the portfolio homogeneity condition (5.1) and exploit that $W_i, \overline{W}_i$ are stochastically independent where $W_i$ follows a GEM distribution with parameters $\tilde{\lambda}_j$ and mixing proportions $\tilde{\pi}_{i,j}$ for $j \in M_{(i)}$, and $\overline{W}_i$ follows a GEM distribution with parameters $\tilde{\lambda}_k$ and mixing proportions $\tilde{\pi}_{\neg i,k}$ for $k \in M_{(\neg i)}$, cf. (5.4).

For $\alpha > 1/a$ we obtain:

$$P(S > \alpha x, U_i > x) = P(\sum_{j=1}^{d} V_j > \alpha x, W_i > x/a)$$

$$= \int_{x/a}^{\alpha x} P(\overline{W}_i > \alpha x - u) f_{W_i}(u) du + \int_{\alpha x}^{\infty} f_{W_i}(u) du$$

$$= \sum_{j \in M_{(i)}} \sum_{k \in M_{(\neg i)}} \tilde{\lambda}_j \tilde{\pi}_{i,j} \tilde{\pi}_{\neg i,k} \exp(-\tilde{\lambda}_k \alpha x) \int_{x/a}^{\alpha x} \exp((\tilde{\lambda}_k - \tilde{\lambda}_j) u) du$$

$$+ \sum_{j \in M_{(i)}} \tilde{\pi}_{i,j} \exp(-\tilde{\lambda}_j \alpha x)$$

$$
\begin{aligned}
&= \sum_{j \in M_{(i)}} \sum_{k \in M_{(\neg i)}} \frac{\tilde{\lambda}_j \tilde{\pi}_{i,j} \tilde{\pi}_{\neg i,k}}{\tilde{\lambda}_k - \tilde{\lambda}_j} \left[ \exp(-\tilde{\lambda}_j \alpha x) - \exp\big(-(\tilde{\lambda}_j/a + (\alpha - 1/a)\tilde{\lambda}_k)x\big) \right] \\
&\qquad\qquad\qquad\qquad\qquad\qquad\qquad\qquad\qquad + \sum_{j \in M_{(i)}} \tilde{\pi}_{i,j} \exp(-\tilde{\lambda}_j \alpha x) \\
&= \sum_{j \in M_{(i)}} \tilde{\pi}_{i,j} \Big(1 + \sum_{k \in M_{(\neg i)}} \frac{\tilde{\lambda}_j \tilde{\pi}_{\neg i,k}}{\tilde{\lambda}_k - \tilde{\lambda}_j}\Big) \exp(-\tilde{\lambda}_j \alpha x) \\
&\qquad - \sum_{j \in M_{(i)}} \sum_{k \in M_{(\neg i)}} \frac{\tilde{\lambda}_j \tilde{\pi}_{i,j} \tilde{\pi}_{\neg i,k}}{\tilde{\lambda}_k - \tilde{\lambda}_j} \exp\big(-(\tilde{\lambda}_j/a + (\alpha - 1/a)\tilde{\lambda}_k)x\big).
\end{aligned}
$$

Applying the mixing proportions property (according to (3.3)):

$$
1 + \sum_{k \in M_{(\neg i)}} \frac{\tilde{\lambda}_j \tilde{\pi}_{\neg i,k}}{\tilde{\lambda}_k - \tilde{\lambda}_j} \;=\; 1 - \sum_{k \in M_{(\neg i)}} \tilde{\pi}_{\neg i \cup j,k} \;=\; \tilde{\pi}_{\neg i \cup j,j}
$$

yields:

$$
\begin{aligned}
&P(S > \alpha x, U_i > x) \\
&= \sum_{j \in M_{(i)}} \tilde{\pi}_{i,j} \exp(-\tilde{\lambda}_j x/a) \Big[ \tilde{\pi}_{\neg i \cup j,j} \exp(-\tilde{\lambda}_j (\alpha - 1/a)x) \\
&\qquad\qquad\qquad\qquad\qquad\qquad - \sum_{k \in M_{(\neg i)}} \frac{\tilde{\lambda}_j \tilde{\pi}_{\neg i,k}}{\tilde{\lambda}_k - \tilde{\lambda}_j} \exp\big(-\tilde{\lambda}_k(\alpha - 1/a)x\big) \Big] \\
&= \sum_{j \in M_{(i)}} \tilde{\pi}_{i,j} \exp(-\tilde{\lambda}_j x/a) \sum_{k \in M_{(\neg i \cup j)}} \tilde{\pi}_{\neg i \cup j,k} \exp\big(-\tilde{\lambda}_k(\alpha - 1/a)x\big) \qquad (7.3) \\
&= \sum_{j \in M_{(i)}} \tilde{\pi}_{i,j} P(V_j > x/a) \, P(\overline{W}_i + V_j > (\alpha - 1/a)x) \qquad\qquad\qquad (7.4)
\end{aligned}
$$

Consequently, we obtain:

$$
P(S > \alpha x \mid U_i > x) \;=\; \frac{\sum_{j \in M_{(i)}} \tilde{\pi}_{i,j} P(V_j > x/a) \, P(\overline{W}_i + V_j > (\alpha - 1/a)x)}{\sum_{j \in M_{(i)}} \tilde{\pi}_{i,j} P(V_j > x/a)}.
$$

Asymptotically for $x \to \infty$ in (7.3) the summands with $j = \tilde{m}(i)$ and $k = \min(\tilde{m}(\neg i), \tilde{m}(i)) = \tilde{m}$ dominate, such that we have:

$$
P(S > \alpha x \mid U_i > x) \;\sim\; \tilde{\pi}_{\neg i \cup \tilde{m}(i), \tilde{m}} \exp(-(\alpha - 1/a)\tilde{\ell}x).
$$

This gives the result for $\alpha > 1/a$; for $\alpha \leq 1/a$ we simply have $P(S > \alpha x, U_i > x) = P(W_i > x/a)$ and, hence, $P(S > \alpha x \mid U_i > x) = 1$. $\qquad\square$